\documentclass[]{spie}  

\usepackage{amsmath,amsfonts,amssymb}
\usepackage{graphicx}
\usepackage[colorlinks=true,allcolors=blue]{hyperref}
\usepackage{orcidlink}

\newcommand{\ket}[1]{\left\lvert #1 \right\rangle}

\title{Scaling up multi-mode entanglement\\ generated by mode swapping}

\author[a]{Yuki Kodama\,\orcidlink{0009-0004-1469-4549}}
\author[a]{Holger F. Hofmann\,\orcidlink{0000-0001-5649-9718}}

\affil[a]{Graduate School of Advanced Science and Engineering,
Hiroshima University, Higashi-Hiroshima 739-8527, Japan}

\authorinfo{Further author information: (Send correspondence to Y.K.)\\
Y.K.: E-mail: d265002@hiroshima-u.ac.jp
\\H.F.H.: E-mail: hofmann@hiroshima-u.ac.jp}

\begin{document}
\maketitle

\begin{abstract}
Entanglement between two local multi-photon multi-mode systems can be generated by swapping a pair of modes between the two local multi-mode systems. In this presentation, we consider possible strategies for the efficient generation of entanglement between systems with three or more modes. It is shown that the choice of photon number inputs in the local systems adds a new degree of freedom to the non-local interference effects observed in the output photon number statistics.
\end{abstract}


\section{INTRODUCTION}
In photonic quantum information processing, entanglement is an essential quantum resource that characterizes non-classical correlations of light, and multi-photon interference plays a crucial role in its generation \cite{pan2012multiphoton,fabre2020modes}. The large-scale implementation of quantum information technologies, including quantum computing \cite{knill2001scheme,kok2007linear,okamoto2009entanglement}, quantum communication\cite{gisin2007quantum}, and quantum metrology\cite{giovannetti2011advances}, requires increasingly large and complex linear-optical circuits capable of manipulating many optical modes and photons. To advance these technologies, it is therefore important to establish a theoretical framework that clarifies how entanglement and photon number correlations are generated in multi-mode multi-photon systems and enables a systematic understanding of the underlying physical mechanisms\cite{Wu_2017,kiyohara2020direct,wiseman2003entanglement}.

In multi-photon interference in a linear optical circuit, the output photon number distribution for a given input photon number state is determined by the linear relationship between the input and output modes. Optical modes are described by creation and annihilation operators, and their input-output relations determine the output photon number distribution. Multi-photon interference is therefore best understood as a consequence of the input output mode transformation. A representative example is the Hong-Ou-Mandel effect\cite{hong1987measurement}. When two photons are incident on a balanced beam splitter, the probability amplitudes for both photons being transmitted and for both photons being reflected cancel each other because of the relative phase imposed by the beam splitter mode transformation. The orthogonality of the input modes ensures that both photons are transmitted to the same output mode.
 
The mode-swapping operation extends the HOM effect to spatially separated multi-mode systems \cite{10.1117/12.3063247,kodama2026non}. In the mode-swapping implementation introduced in previous work, one spatial mode from each local system is exchanged between the two systems. The input modes can then be expanded in terms of two non-local modes, each defined as a linear combination of spatially separated modes. As a result, non-local two photon interference and photon number correlations arise between the two spatially separated systems.

Here, we extend the mode swapping operation to a pair of three mode systems, denoted as systems A and B, and analyze how mode swapping affects the output photon number distribution. A local unitary transformation is applied to each system before mode swapping, followed by the corresponding inverse transformation. In the absence of mode swapping, the two local transformation cancel each other. Therefore, any change in the output relative to the input state can be directly attributed to the mode swapping operation.

 In Section~\ref{sec:mode-swapping}, we define the swapped modes as linear combinations of the local modes and show how mode swapping connects each local output mode to modes in the other system. In Section~\ref{sec:analysis}, we calculate the output state and the photon number distribution for the specific asymmetric three photon input state $\ket{200}_A\ket{100}_B$ and discuss the multiphoton interference and the correlations between two system generated by mode swapping.

\begin{figure}[htbp]
\vspace{0.5cm}
    \centering
    \includegraphics[width=0.70\linewidth]{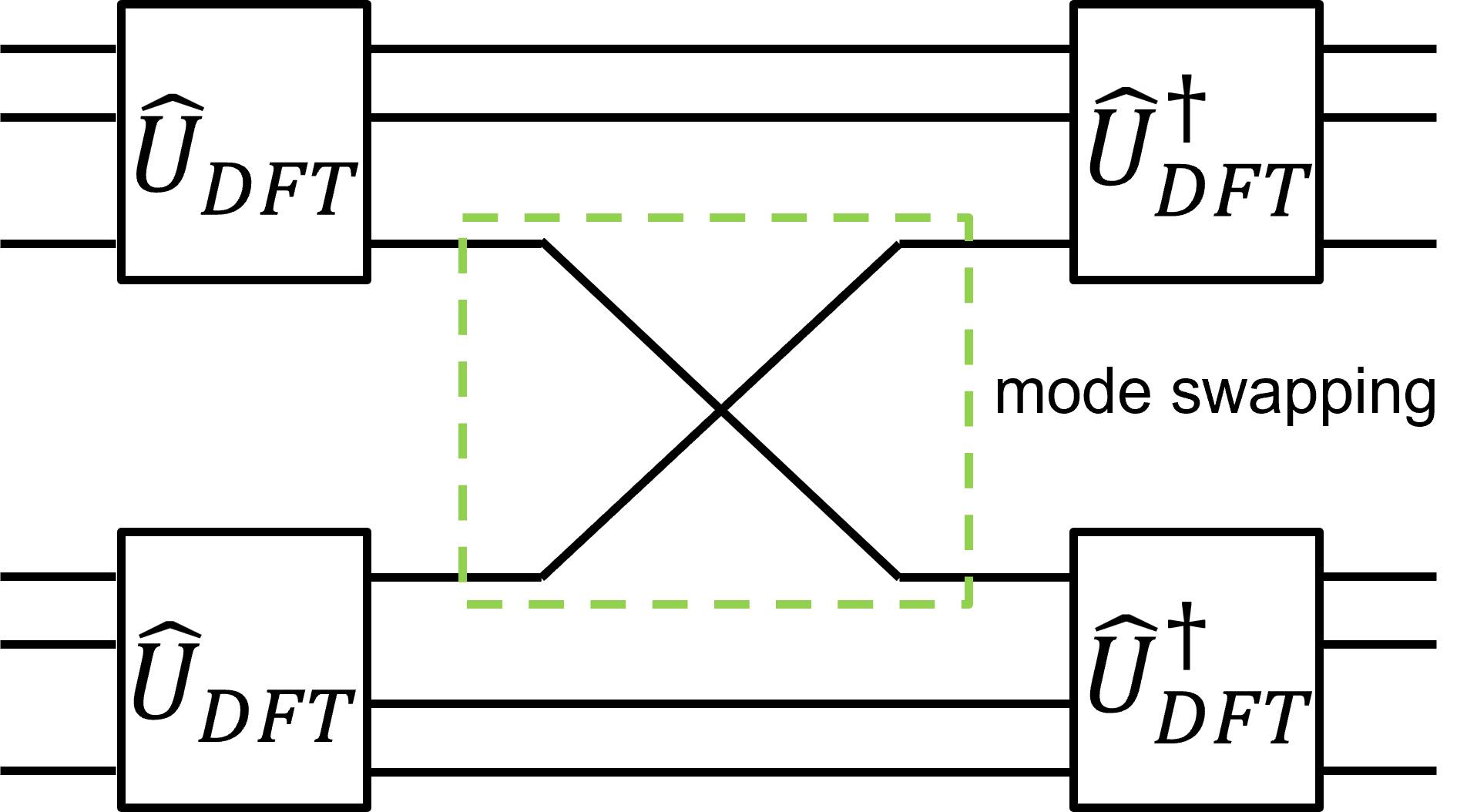}
\vspace{0.5cm}
    \caption{Schematic representation of mode swapping between two three-mode interferometers. A local discrete Fourier transform (DFT) is applied to the modes to ensure that an equal superposition of input modes is exchanged between systems $A$ and $B$. The inverse DFT is applied afterwards to identify the output ports with the local input modes.}
    \label{fig:circuit1}
\end{figure}

\section{OPTICAL SETUP AND MODE-SWAPPING TRANSFORMATION}
\label{sec:mode-swapping}
Figure~\ref{fig:circuit1} shows the mode-swapping interferometer considered in this work. The interferometer consists of two systems, $A$ and $B$, each containing three local modes. A local unitary transformation is applied to each system before the mode-swapping operation, and the inverse transformation is applied after the swap. Therefore, in the absence of mode swapping, the two local transformations cancel each other, and the input photon state appears unchanged at the output.

When one pair of modes is exchanged in the intermediate paths, however, non-local correlations are introduced between the two systems. As a result, the output state generally exhibits a photon number distribution different from that of the input state, and entanglement is generated between systems $A$ and $B$ for appropriate multi-photon inputs. Since the difference between the input and output states is produced by a single mode-swapping operation, this circuit provides a direct way to investigate the correlations generated between the two systems by mode exchange.
The creation operators of the three modes are denoted by $\{\hat a_m^\dagger\}_{m=0}^{2}$ and $\{\hat b_m^\dagger\}_{m=0}^{2}$
for systems $A$ and $B$, respectively.
In the present setup, the local unitary transformations are chosen such that the modes $\hat s_A$ and $\hat s_B$ exchanged in the intermediate paths correspond to unbiased superpositions of the input modes of the respective local systems. This relation can be realized by applying the following discrete Fourier transform (DFT),
\begin{equation}
\hat U_{\mathrm{DFT}}
\hat a_k^\dagger
\hat U_{\mathrm{DFT}}^\dagger
=
\sum_{m=0}^{M-1}
\frac{1}{\sqrt{M}}
\exp\left(i\frac{2\pi}{M}km\right)
\hat a_m^\dagger.
\label{eq:dft_transformation}
\end{equation}
Here the transformation has been written for the modes of system $A$. The same relation holds for the modes $\hat b_k$ of system $B$. In the present setup, each local system consists of three modes, and therefore $M=3$.
As mentioned above, the DFT ensures that the modes $\hat s_A$ and $\hat s_B$ exchanged in the intermediate paths are equal superpositions of the input modes of each local system. If the swapped modes are the $k=0$ outputs of the DFT, they can be expressed as
\begin{align}
\hat s_A^\dagger
&=
\frac{1}{\sqrt{M}}
\sum_{m=0}^{M-1}\hat a_m^\dagger,
&
\hat s_B^\dagger
&=
\frac{1}{\sqrt{M}}
\sum_{m=0}^{M-1}\hat b_m^\dagger.
\label{eq:swapped_mode_definition}
\end{align}
Let $\hat U_{\mathrm{swap}}$ denote the unitary describing the swap operation. This unitary acts only on the creation operators of the two exchange ports and leaves all other intermediate modes unchanged. The swap operation acts only on the swapped modes and has no effect on the remaining mode components. Therefore, each creation operator $\hat a_m^\dagger$ can be decomposed into a component within the swapped-mode subspace and a component orthogonal to it. Since the overlap between $\hat a_m$ and $\hat s_A$ is determined by $[\hat a_m,\hat s_A^\dagger]=\frac{1}{\sqrt{M}}$,
the creation operator can be written as
\begin{equation}
\hat a_m^\dagger
=
\left(
\hat a_m^\dagger
-
\frac{1}{\sqrt{M}}\hat s_A^\dagger
\right)
+
\frac{1}{\sqrt{M}}\hat s_A^\dagger.
\label{eq:mode_decomposition}
\end{equation}
The first term is orthogonal to $\hat s_A$. Since this component is unaffected by the central swap operation, it remains unchanged under the complete DFT--swap--inverse-DFT transformation. In contrast, $\hat s_A^\dagger$ is transformed into $\hat s_B^\dagger$ by the swap operation. Consequently,
\begin{equation}
\hat U_{\mathrm{swap}}
\hat a_m^\dagger
\hat U_{\mathrm{swap}}^\dagger
=
\hat a_m^\dagger
+
\frac{1}{\sqrt{M}}
\left(
\hat s_B^\dagger-\hat s_A^\dagger
\right).
\label{eq:A_mode_transform}
\end{equation}
Similarly, for the modes of system $B$,
\begin{equation}
\hat U_{\mathrm{swap}}
\hat b_m^\dagger
\hat U_{\mathrm{swap}}^\dagger
=
\hat b_m^\dagger
-
\frac{1}{\sqrt{M}}
\left(
\hat s_B^\dagger-\hat s_A^\dagger
\right).
\label{eq:B_mode_transform}
\end{equation}
These results show that the action of mode swapping is restricted to the two-mode subspace $\operatorname{span}\{\hat s_A,\hat s_B\}$ spanned by the two swapped modes. The effect of mode swapping therefore appears through the difference between the two swapped-mode operators, $\hat s_B^\dagger-\hat s_A^\dagger$.

\section{ANALYSIS OF THE POSTSELECTED THREE-PHOTON STATE}
\label{sec:analysis}
In the previous section, we showed how the mode swapping operation transforms the optical modes. We can now define a number of input scenarios by selecting the input photon numbers of each input mode $\hat{a}_m$ and $\hat{b}_m$. Here, we select an asymmetric three-photon input state,
\begin{equation}
\ket{\Psi_{\mathrm{in}}}
=
\ket{200}_A\ket{100}_B
=
\frac{1}{\sqrt{2}}
\left(\hat a_0^\dagger\right)^2
\hat b_0^\dagger
\ket{\mathrm{vac}}.
\label{eq:input_state}
\end{equation}
In this input state, two photons are injected into mode $0$ of system $A$, while one photon is injected into mode $0$ of system $B$. The final output state is obtained by applying the swap unitary operation $\hat U_{\mathrm{swap}}$ to the input state. Using Eq.~(\ref{eq:A_mode_transform}) and (\ref{eq:B_mode_transform}), the state after passing through the interferometer is
\begin{align}
\ket{\Psi_{\mathrm{MS}}}
=
\frac{1}{\sqrt{2}}
\left[
\hat a_0^\dagger
+
\frac{1}{\sqrt{3}}
\left(
\hat s_B^\dagger-\hat s_A^\dagger
\right)
\right]^2
\left[
\hat b_0^\dagger
-
\frac{1}{\sqrt{3}}
\left(
\hat s_B^\dagger-\hat s_A^\dagger
\right)
\right]
\ket{\mathrm{vac}}.
\label{eq:full_swapped_state}
\end{align}
Since mode swapping transfers photons between the two local systems, this state contains not only components with two photons in system $A$ and one photon in system $B$, but also components with different local photon numbers. The initial local photon number distribution $(N_A,N_B)=(2,1)$ is obtained when the changes in the photon numbers contained in the swapped modes of the two systems cancel each other.
As shown in Eq.~(\ref{eq:mode_decomposition}), each creation operator can be decomposed into a component contained in the swapped mode and a component orthogonal to it that is unaffected by mode swapping. Therefore, the local photon number distribution $(2,1)$ is preserved either when all three photons are contained in components orthogonal to the swapped modes, so that no photon transfer occurs between the systems, or when one photon is exchanged in each direction between systems $A$ and $B$.
Let $\hat\Pi_{2,1}$ denote the projector onto the subspace with local photon numbers $(2,1)$. After the post-selection, only the term without photon transfer and the term describing the reciprocal transfer of one photon in each direction remain. Therefore, the state after the post-selection is given by
\begin{align}
\ket{\Psi_{2,1}^{\mathrm{(u)}}}
&:=
\hat\Pi_{2,1}\ket{\Psi_{\mathrm{MS}}}
\nonumber\\
&=
\frac{1}{\sqrt{2}}
\left(
\hat a_0^\dagger-\frac{1}{\sqrt{3}}\hat s_A^\dagger
\right)^2
\left(
\hat b_0^\dagger-\frac{1}{\sqrt{3}}\hat s_B^\dagger
\right)
\ket{\mathrm{vac}}
\nonumber\\
&\quad+
\frac{2}{3\sqrt{2}}
\left(
\hat a_0^\dagger-\frac{1}{\sqrt{3}}\hat s_A^\dagger
\right)
\hat s_A^\dagger
\hat s_B^\dagger
\ket{\mathrm{vac}}.
\label{eq:unnormalized_postselected_state}
\end{align}
The first term corresponds to the case in which the photon numbers in the swapped modes are zero. The second term corresponds to the case in which the photon numbers in each swapped mode is one and the two photons are exchanged reciprocally between systems $A$ and $B$.
The success probability of this post-selection is given by the sum of the probabilities of these two processes. The respective probabilities are
\begin{equation}
P_{\mathrm{no-swap}}
=
\frac{8}{27},
\qquad
P_{\mathrm{swap}}
=
\frac{4}{27}.
\label{eq:path_probabilities}
\end{equation}
Therefore, the success probability of the $(2,1)$ post-selection is
\begin{equation}
P_{2,1}
=
\frac{8}{27}
+
\frac{4}{27}
=
\frac{4}{9}
.
\label{eq:postselection_probability}
\end{equation}
Although mode swapping is performed on only one of the three local modes, the probability that the photon number in each system remains equal to its initial value is smaller than one half.

After normalizing the postselected state and using Eq.~(\ref{eq:swapped_mode_definition}) to expand it in the Fock basis of the external modes, we obtain
\begin{align}
\ket{\Psi_{2,1}}
={}&
\frac{1}{6}
\left(
4\ket{200}_A
-\sqrt{2}\ket{110}_A
-\sqrt{2}\ket{101}_A
\right)\ket{100}_B
\nonumber\\
&-
\frac{1}{6}
\left(
\ket{020}_A
+\ket{002}_A
-\sqrt{2}\ket{110}_A
-\sqrt{2}\ket{101}_A
+\sqrt{2}\ket{011}_A
\right)
\nonumber\\
&\hspace{20mm}\times
\left(
\ket{010}_B+\ket{001}_B
\right).
\label{eq:final_fock_state}
\end{align}

Table~\ref{tab:conditional_probabilities} shows the joint photon number probabilities conditioned on the $(2,1)$ post-selection. The probabilities before post-selection can be obtained by multiplying each entry in the table by the post-selection success probability $P_{2,1}=4/9$. 
\begin{table}[htbp]
\centering
\caption{Joint probabilities $P({\bf n}_A,{\bf n}_B)$ for the input state $\ket{200}_A\ket{100}_B$. According to the post-selection condition, ${\bf n}_A$ is one of six possible distributions of two photons among the three modes of system $A$ and ${\bf n}_B$ is one of three possible distributions of one photon among the three modes of system $B$.}
\vspace{0.5cm}
\label{tab:conditional_probabilities}
\renewcommand{\arraystretch}{1.25}
\begin{tabular}{c|cccccc}
${\bf n}_B \backslash {\bf n}_A$
& $200$ & $020$ & $002$ & $110$ & $101$ & $011$ \\
\hline
$100$ & $16/36$ & $0$ & $0$ & $2/36$ & $2/36$ & $0$ \\
$010$ & $0$ & $1/36$ & $1/36$ & $2/36$ & $2/36$ & $2/36$ \\
$001$ & $0$ & $1/36$ & $1/36$ & $2/36$ & $2/36$ & $2/36$
\end{tabular}
\end{table}
As a result of a single mode swap of the unbiased superposition of all input modes, the conditional probability that the output photon number distribution is the same as the input distribution $\ket{200}_A\ket{100}_B$ is $4/9$, where only results with $N_A=2$ and $N_B=1$ are considered. Although larger than one half in total, the individual probabilities of all other output photon number distributions are significantly smaller than the probability of obtaining the original distribution $\ket{200}_A\ket{100}_B$, indicating a certain degree of randomness in the re-distribution of photons caused by the mode swap. However, the outcomes with zero probability indicate that the distribution of outcomes in Table~\ref{tab:conditional_probabilities} result from coherent interference between the probability amplitudes associated with zero phtons in the swapped modes and with one photon each in both of the swapped modes. It might be worth noting that the distribution is symmetric under the exchange of input modes 1 and 2 in each system. This is because only input mode $0$ is occupied in each system in the input state, while the swapped modes are equal superpositions of all input modes, as shown in Eq.~(\ref{eq:swapped_mode_definition}). Consequently, modes 1 and 2 are equivalent.

When both photons in system $A$ are detected in mode $0$, the photon in system $B$ is necessarily detected in mode $0$. The zero probabilities of $\ket{200}_A\ket{010}_B$ and $\ket{200}_A\ket{001}_B$ indicate that complete destructive interference occurs between the no-swap and swap amplitudes. Similarly, when the photon in system $B$ is detected in mode $0$, there is no possibility of detecting both photons in mode $1$ or $2$ - the photon number in mode $0$ is either one or two, never zero. These correlations arise from the entanglement between the one photon output in $B$ and the two photon output in $A$ generated by swapping the symmetric modes. This is the most simple scenario of an entangled state between two different photon numbers in the two three-mode systems generated by a single mode swap operation.

\section{CONCLUSIONS}
We extended the mode swapping operation to two spatially separated three-mode interferometers. The modes to be exchanged were defined as equal superpositions of the input modes of the respective local systems, allowing us to derive the transformation of the local creation operators. Any multi-photon input can then be expressed by the corresponding application of these creation operators to the vacuum state. Since the action of mode swapping is confined to the subspace of the two swapped modes, the mode transformations take a particularly simple form. However, swapping equal superpositions of the local input modes ensures that all input modes participate in non-local interference effects. 

A wide variety of input states are possible, and most of them have highly non-trivial characteristics. Here, we demonstrate the generation of entanglement using an asymmetric three-photon input state $\ket{200}_A\ket{100}_B$. By post-selecting only the outputs with two photons in $A$ and one in $B$, we obtain entanglement between one photon is three possible states and two photons in six possible states. If the single photon remains in its original input mode, at least one of the two photons in $A$ will be found in their original mode as well. If the single photon is found in a different mode from its original input mode, at least one of the two photons in $A$ will also have moved to a different mode. The multi-photon interference effects associated with mode swapping thus introduce strong correlations in the output photon number distributions between the input modes of the two local systems. In particular, the probabilities of zero predicted for a large number of possible outcomes illustrates the potential of entanglement generation by mode swapping between two multi-mode multi-photon systems.

The theoretical framework developed here provides the basis for extending the analysis to optical quantum circuits with larger numbers of modes and photons and is expected to contribute to the application of multi-mode multi-photon interference in optical quantum computing and quantum communication.

\acknowledgments
This work was supported by ERATO, Japan Science and Technology Agency (JPMJER2402), and by JST SPRING, Grant Number JPMJSP2132.

\bibliographystyle{spiebib}
\bibliography{report}

\end{document}